# Efficient Collection of Connected Vehicle Data based on Compressive Sensing*

Lei Lin and Srinivas Peeta

***Abstract*— Connected vehicles (CVs) can capture and transmit detailed data like vehicle position, speed and so on through vehicle-to-vehicle and vehicle-to-infrastructure communications. The wealth of CV data provides new opportunities to improve the safety, mobility and sustainability of transportation systems. However, the potential data explosion likely will overburden storage and communication systems. To solve this issue, we design a real-time compressive sensing (CS) approach which allows CVs to collect and compress data in real-time and can recover the original data accurately and efficiently when it is necessary. The CS approach is applied to recapture 10 million CV Basic Safety Message speed samples from the Safety Pilot Model Deployment program. With a compression ratio of 0.2, it is found that the CS approach can recover the original speed data with the root mean-squared error as low as 0.05. The recovery performances of the CS approach are further explored by time-of-day and acceleration. The results show that the CS approach performs better in data recovery when CV speeds are steady or changing smoothly.**

## I. Introduction

A connected vehicle (CV) is designed to record and exchange its activity data through vehicle-to-vehicle (V2V) and vehicle-to-infrastructure (V2I) communications. With the popularization of CVs, the safety and mobility of the whole transportation system are expected to be improved through enhanced situational awareness and prediction. In 2012, the Safety Pilot Model Deployment (SPMD) program was launched in Ann Arbor, Michigan. Nearly 3,000 vehicles were equipped with GPS antennas and DSRC (Dedicated Short-Range Communications) devices, each of which broadcasts the Basic Safety Message (BSM) that contains the position, velocity, and yaw rate, to neighboring CVs and nearby roadside units (RSUs) at a rate of 10 Hz [1]. These CV data provide opportunities for intelligent transportation system applications like accurate traffic state estimation and prediction [2], [3], traffic signal optimization [4], etc.

However, the collection of detailed CV data would likely overburden current storage and communication systems. One report suggests that one CV will upload 25 GB data to the cloud every hour [5]. Muckell et al. (2014) provide a specific example on the communication cost. Assuming that transmitting one megabyte data over cellular or satellite networks costs $5 to $7 typically, tracking a fleet of 4,000 vehicles would cost approximately $1.8-$2.5 million annually [6]. The storage and communication costs would be considerable as the market penetration rate (MPR) of CVs increases.

To address the cost challenges, previous studies have focused on two approaches. The first approach is called sample-then-compression, which collects data at a fixed rate first and compresses the data offline. Muckell et al. (2014) present a GPS trajectory compression method called SQUISH-E (Spatial QUalIty Simplification Heuristic - Extended) [6]. It can acquire the optimal compression ratio with a user-specified error bound. The limitation of the approach is that it lacks optimal solutions to control data collection rates. Hence, redundant or unnecessary data are captured, transmitted and stored.

By contrast, the second approach develops dynamic data capture strategies to reduce the total amount of vehicle-based data and satisfies the system awareness and control needs of transportation authorities. It uses the concept of Dynamic Interrogative Data Capture (DIDC) [7]. Its basic idea is to identify the smallest data collection and transmission rates to provide information requested by the transportation authority on different traffic situations (for example, system-wide travel time in the vicinity of an accident, or travel time on a link under normal traffic conditions). When multiple requests are received, a DIDC controller conducts a heuristic optimization routine to prioritize the most important one. However, the main issue with the second approach is that the requests received by the DIDC controller may be conflicting with each other and the prioritization and sorting of the requests are not trivial.

In this study, for the first time, we design a compressive sensing (CS) approach for CV data collection. CS has become an active research area in recent years as a revolutionary new approach to capture a wide range of signals, mainly due to several recently-obtained important results [8], [9]. CS is different from the first approach which first acquires huge amounts of data and then discards a significant portion of redundant data in the compression stage [10]. Instead, it is a real-time compression approach which enables redundancy in the signal removed during the sampling process, leading to a lower but more effective sampling rate [11]. Further, unlike the second approach, CS does not adapt dynamic data collection rates based on various performance measurement requests [4]. Essentially, it conducts a non-adaptive linear transform that relies on two well-designed matrices, sensing

*This work is supported by the NEXTRANS Center at Purdue University, and CCAT, the Region 5 University Transportation Center.

L. Lin is with the NEXTRANS Center, Purdue University, West Lafayette, IN 47906 USA (e-mail: lin954@purdue.edu).

S. Peeta is with School of Civil Engineering, Purdue University, West Lafayette, IN 47906 USA, and also with the NEXTRANS Center, Purdue University, West Lafayette, IN 47906 USA (e-mail: peeta@purdue.edu).

matrix and sparsifying matrix, to capture the essential information embedded in the original signal with a very low sampling rate [12]. The original signal can still be recovered with high accuracy. Therefore, there is no need to deal with multiple conflicting data requests.

As will be shown later, the proposed CS approach is very straightforward and can be easily implemented with current CV data collection approaches. Specifically, a CV still collects data samples at a fixed rate, except now for every sample we have the CS approach to determine whether to keep it or not. Furthermore, the approach allows the recovery of CV data with high accuracy when necessary. We use the CS approach to collect 10 million real-world CV speed samples from the SPMD program. The results show that a well-designed CS approach can greatly reduce the amount of CV data with a compression ratio of 0.2, and the original CV speed data can be recovered with a root mean-squared error as low as 0.05.

The rest of the paper is organized as follows. The next section introduces basics of CS theory and the CS approach for CV data collection. Then, the CS approach is applied to recapture the information in the 10 million BSM speed samples, and the accuracy of the recaptured data is analyzed by time-of-day and acceleration. The paper concludes with a discussion of the study findings and future research directions.

## II. Methodology

This section first introduces the basic concepts of CS theory. Then, it presents the CS approach for CV data collection and recovery.

### A. *Compressive Sensing*

Consider a signal vector $\mathrm{x} \in R^N$. It can be represented in terms of a set of orthonormal bases $\{\Psi_i\}_i^N$, $\Psi_i \in R^N$ as:

$$\mathrm{x} = \Psi\alpha \qquad (1)$$

The signal x is $K$-sparse if α, the transformed coefficient vector, has $K$ nonzero entries. Ψ is an $N \times N$ matrix called sparsifying matrix.

In the traditional sample-then-compression approach, the full signal vector x is acquired first, and then the vector α is computed through $\alpha = \Psi^T \mathrm{x}$, by keeping only the $K$ largest coefficients and discarding the rest [13]. As discussed earlier in the introduction section, this approach is not efficient, because the original CV data are still collected, transmitted and stored before being compressed.

By contrast, CS directly acquires a compressed signal through the following sampling process:

$$\mathrm{y} = \Phi\mathrm{x} = \Phi\Psi\alpha = \Theta\alpha \qquad (2)$$

where $\mathrm{y} \in R^M$ is the sampled vector, $M \ll N$, Φ is an $M \times N$ matrix called the measurement matrix, and $\Theta = \Phi\Psi$ is an $M \times N$ matrix. The ratio $M/N$ is labeled the compression ratio.

Equation (2) defines an underdetermined linear system, because the number of equations $M$ is much less than the number of unknown entries $N$ [11]. With prior knowledge of the sparsity, α can be recovered from y which consists of $M$ measurements by solving the following $l_0$ norm minimization problem:

$$\arg min_\alpha \|\alpha\|_0, \text{ subject to } \Theta\alpha = \mathrm{y} \qquad (3)$$

where the $l_0$ norm $\|\alpha\|_0$ is the number of non-zero elements in the vector, which measures the signal sparsity.

After α is available, the original signal estimation $\hat{\mathrm{x}}$ can be calculated using Equation (1). However, this is an NP-hard problem with no efficient solutions. To solve this issue, the CS theory introduces the following definition [11]: matrix A satisfies the restricted isometry property (RIP) of order $K$, if there exists a constant $\delta_K \in (0, 1)$ such that:

$$(1 - \delta_K)\|\mathrm{v}\|_2^2 \leq \|\mathrm{Av}\|_2^2 \leq (1 + \delta_K)\|\mathrm{v}\|_2^2 \qquad (4)$$

for all v satisfying $\|\mathrm{v}\|_0 \leq K$.

It has been proved that if matrix Θ satisfies the RIP of order $2K$, the following $l_1$ norm optimization problem can be solved to obtain an accurate reconstruction [11]:

$$\arg min_\alpha \|\alpha\|_1, \text{ subject to } \Theta\alpha = \mathrm{y} \qquad (5)$$

Because Θ satisfies the RIP of order $2K$ as shown in Equation (6), it implies that the distance between any pair of $K$-sparse signals $\alpha_1$ and $\alpha_2$ does not shrink or extend too much due to the dimensionality reduction from $\alpha \in R^N$ (original data) to $\mathrm{y} \in R^M$(sampled data). That is, the salient information in any $K$-sparse signal is not damaged [13].

$$(1 - \delta_{2K})\|\mathrm{v}\|_2^2 \leq \|\Theta\mathrm{v}\|_2^2 \leq (1 + \delta_{2K})\|\mathrm{v}\|_2^2 \qquad (6)$$

where $\mathrm{v} = \alpha_1 - \alpha_2$ and $\|\mathrm{v}\|_0 \leq 2K$.

Linear programming and other convex optimization algorithms have been proposed to efficiently solve Equation (5).

### B. *The CS Approach for CV Data Collection and Recovery*

Suppose $\mathrm{x} \in R^N$ is a vector of CV data samples, e.g., speed samples collected at a fixed rate. According to Equation (1), we need to conduct a transform $\alpha = \Psi^T\mathrm{x}$ so that α has sparse coefficients in that domain. Typical transforms in the literature include discrete cosine transform (DCT) and Discrete Wavelet Transform (DWT). DCT is a Fourier-related transform like discrete Fourier transform (DFT), but uses cosine functions and the transformed coefficients are real numbers. DWT is more suitable for piecewise constant signals [11], and is not applicable to fluctuating speed samples. Therefore, DCT is chosen in this study. More specifically it is shown that [14]:

$$\alpha_j = K(j) \sum_{i=1}^{N} \mathrm{x}_i cos\frac{\pi j(i-0.5)}{N}, j = 0, \ldots, N-1 \qquad (7)$$

where $K(j) = \frac{1}{\sqrt{N}}$ when $j = 0$;

$K(j) = \sqrt{\frac{2}{N}}$ when $1 \leq j \leq N-1$.

Next, we need to select a matrix $\Theta$ to obtain the sampled vector $y \in R^M$ based on Equation (2). $\Theta$ should satisfy the RIP of order $2K$ so the original vector x can be recovered. Previous studies have proved the following theorem:

**Theorem 1** [15] Suppose a $M \times N$ matrix $\Theta$ is obtained by selecting $M$ rows independently and uniformly at random from the rows of a $N \times N$ unitary matrix U. Normalize the matrix $\Theta$ so that its columns have unit $l_2$ norms. Then, $\Theta$ satisfies the RIP with high probability $1 - N^{-O(\delta_{2K}^2)}$ for every $\delta_{2K} \in (0, 1)$ provided that

$$M = \Omega(\mu_U^2 K log^5 N) \quad (8)$$

where $\mu_U = \sqrt{N} max_{i,j}|u_{i,j}|$ is called the coherence of the unitary matrix U.

Following Theorem 1, we select an $N \times N$ inverse discrete cosine transform (IDCT) matrix $\Psi$ as the unitary matrix U, and randomly select $M$ rows out of it to build the matrix $\Theta$. The main advantage is that it allows us to skip the DCT and IDCT transforms and acquire $M$ samples y directly from the real observations x due to the following:

$$y = \Theta\alpha = D\Psi\Psi^T x = Dx \quad (9)$$

where $\Theta = D\Psi$; D is a random subset of $M$ rows of the identity matrix of size $N \times N$. If another matrix were chosen as U instead of the IDCT matrix $\Psi$, it would need matrix calculations to determine $\alpha$ and y.

Finally, with the determination of matrix $\Psi$ and $\Theta$, the CS approach for CV data collection and recovery can be summarized as follows: Suppose a single CV is collecting speed samples at a fixed rate. A speed sample is kept if the randomly generated value from a uniform distribution over $[0, 1]$ is less than or equal to the compression ratio $M/N$; otherwise it is discarded. In the future, when the original data is needed for certain applications, it can be reconstructed by solving the $l_1$ norm optimization problem defined in Equation (5) to reconstruct $\alpha$, which can then be converted back to obtain the recovered data $\hat{x}$. The following case study illustrates the application of the proposed CS approach in detail.

## III. Collection and Recovery of 10 Million BSM Speed Samples

In the SPMD program, BSMs are generated for each CV at 10 Hz according to the SAE J2735 standard [1]. The basic components of the BSMs include the device ID, time stamp, latitude, longitude, vehicle speed, vehicle heading, yaw rate, and radius of curve [16]. We extract 10 million BSM speed samples from this dataset. The basic statistics of the 10 million speed samples are listed in Table 1.

TABLE I. Basic Statistics of 10 Million BSM Speed Samples

| | |
|---|---|
| Time Period | 04/01/2013 – 04/03/2013 |
| Number of CVs | about 3,000 |
| Number of Trips | 16,798 |
| Mean (m/s) | 17.23 |
| Standard Deviation (m/s) | 10.83 |

As shown in Table 1, the BSM speed samples are collected from nearly 3,000 CVs in Ann Arbor, Michigan from 04/01/2013 to 04/03/2013. To protect the privacy of the SPMD participants, the BSMs that either contain Personally Identifiable Information (PII) or may lead to the discovery of PII are removed, such as the origin and destination of a participant. Further, there are no Trip IDs included in the raw data; hence, we convert the 10 million BSM speed samples into 16,798 trips by checking whether the timestamps are continuous, e.g., the gap should be 0.1 seconds with the fixed rate of 10 Hz. The mean and standard deviation of the speed data are 17.23 m/s and 10.83 m/s.

Before applying the proposed CS approach for CV data capture, sparsity analysis is conducted to check whether the original speed samples are sparse with the DCT. As an illustration, Fig. 1 shows a set of BSM speed samples x with $N = 1000$ and the corresponding DCT coefficients $\alpha$. It can be observed that $\alpha$ is a sparse vector and only about 50 coefficients have non-zero values.

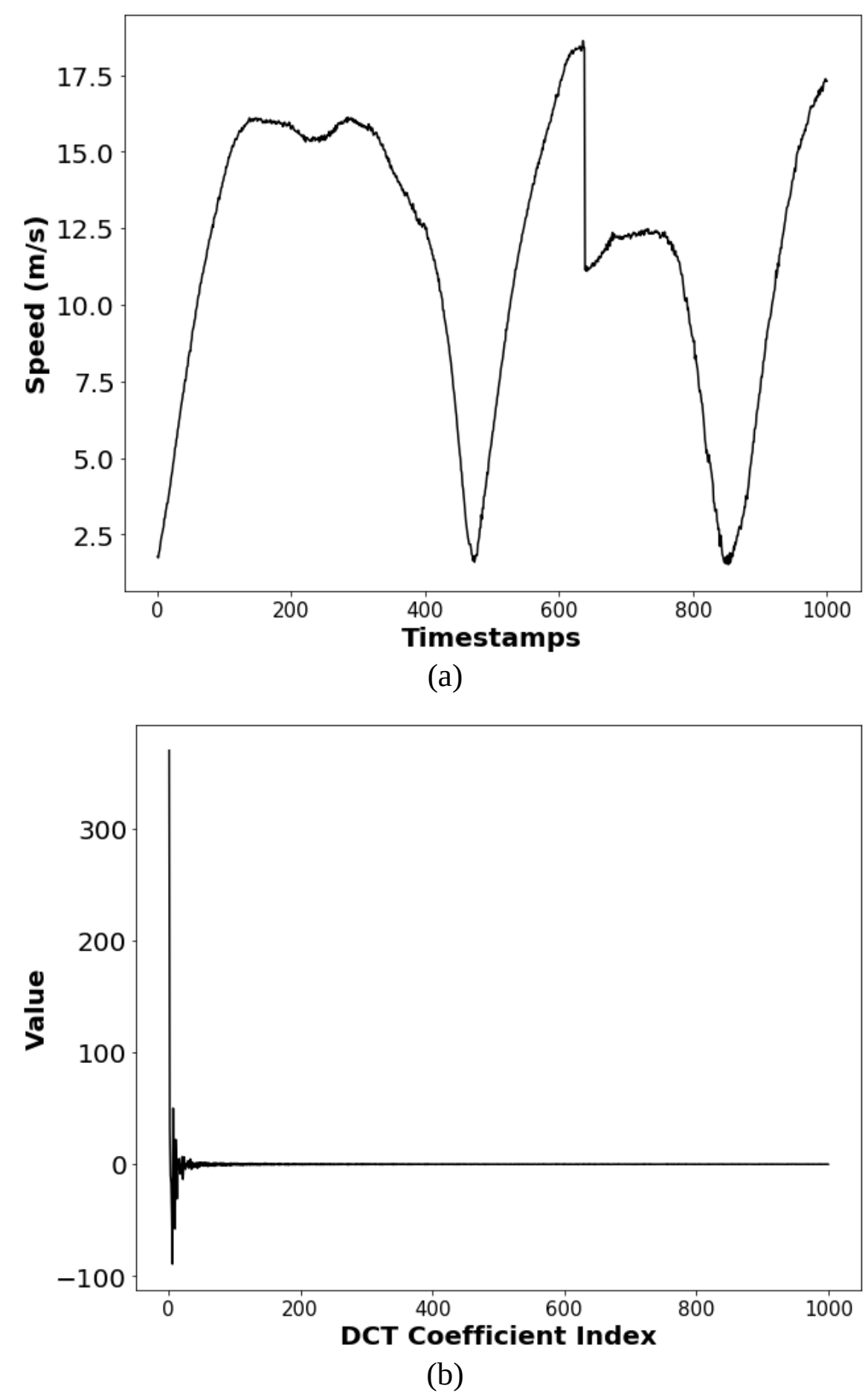

Fig. 1. Sparsity Analysis (*a. the Original 1,000 BSM Speed Samples; b. the 1,000 DCT Coefficients*)

After the sparsity verification, the proposed CS approach for CV data collection is implemented. As introduced in the methodology section, the CV collects speed samples at a fixed rate. We determine the compression ratio $M/N$, and keep or discard each speed sample by drawing a value from a uniform distribution over $[0, 1]$; the sample is kept if the value is less than or equal to the compression ratio, and discarded otherwise.

When the original CV speed data needs to be recovered for a certain application, e.g., travel time estimation, for every $M$ measures, the $l_1$ norm optimization in Equation (5) is solved to convert $y \in R^M$ to the coefficients in DCT domain $\alpha \in R^N$. The IDCT is then performed on α to recover the original data $\hat{x} \in R^N$. If with the CS approach the length of the compressed CV dataset is $L$, then to recover the original data, we need to solve Equation (5) $L/M$ times. Note that the whole recovery process can be easily implemented in parallel. The recovery accuracy is measured using the root mean squared error (RMSE) value normalized with respect to the $l_2$ norm of the whole series of data [11]:

$$\text{RMSE} = \frac{\|x_o - \hat{x}_r\|_2}{\|x_o\|_2} \quad (10)$$

where $x_o$ is the original 10 million speed samples, and $\hat{x}_r$ is the recovered BSM speed data.

$M$ and $N$ are two parameters in the CS approach that determine the compression ratio and affect the recovery accuracy. Fig. 2a shows the RMSEs for the whole 10 million speed samples by compression ratio with different settings of $M$ and $N$. Furthermore, the computational complexity of the $l_1$ norm optimization in Equation (6) is about $O(N^3 + MN^2)$ [13]. The average time consumption per recovery is also calculated for each setting and shown in Fig. 2b. All experiments are conducted in Windows 10, i7-6820HK CPU with 64 GB RAM.

As can be seen in Fig. 2a, when the compression ratio is 0.1, the RMSE with the setting of $N = 1000$ is lower than under all the other settings. As compression ratio increases, the RMSEs decrease and become close to each other for different values of $N$. When compression ratio is higher than 0.2, the RMSEs are lower than 0.025, and when the compression ratio is 0.6, the RMSEs are almost zero.

As shown in Fig. 2b, the time consumption per recovery is almost zero for all compression ratios when $N = 100$ and $N = 200$. When $N = 500$ and $N = 1000$, the curves of time consumption per recovery increase with larger compression ratios. In particular, time consumed is higher than 1.4 seconds per recovery when the compression ratio is equal to 0.6 and $N = 1000$.

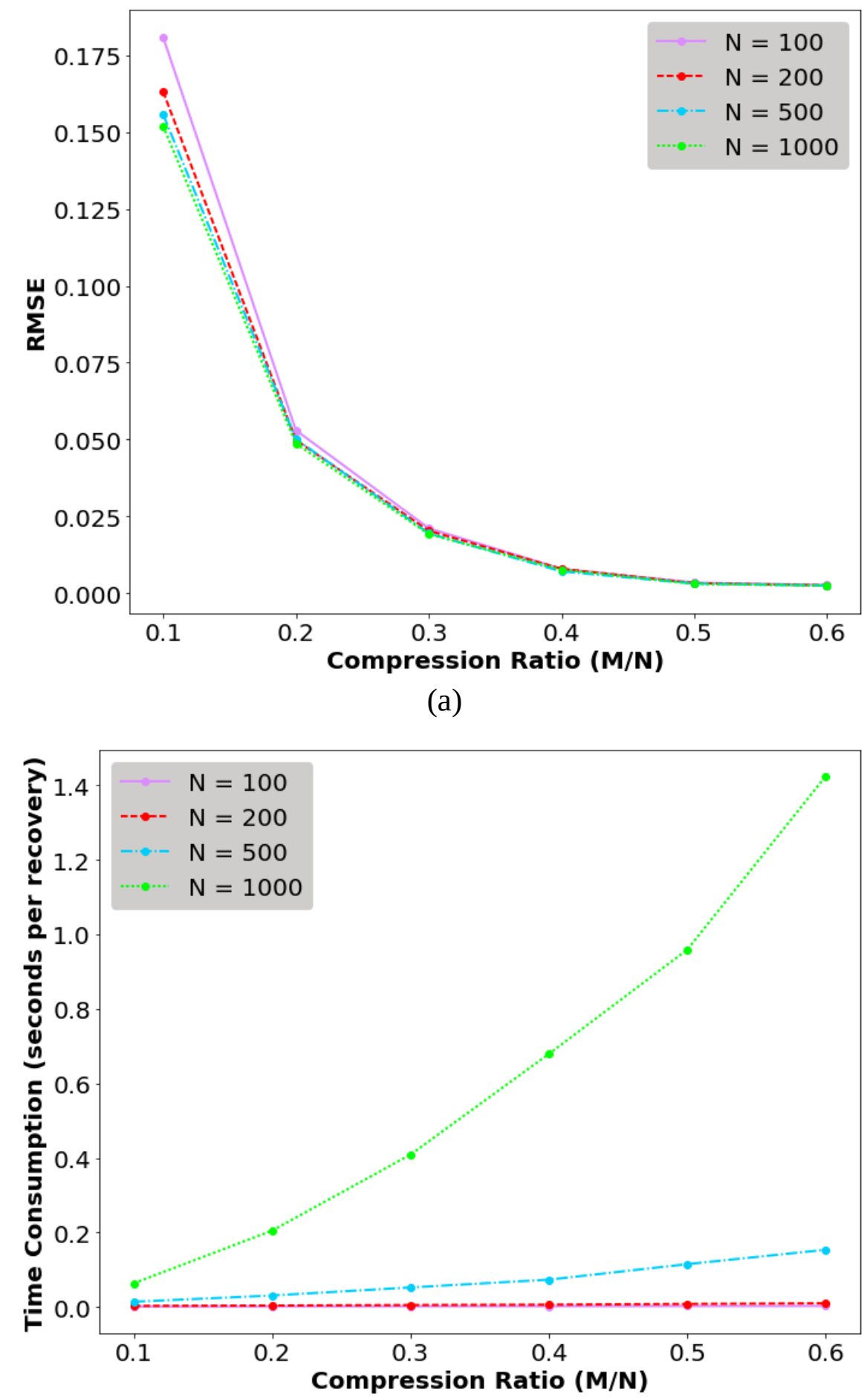

(a)

(b)

Fig. 2. Recovery Performance with Various Settings of M and N (*a. RMSE by Compression Ratio; b. Time Consumption per Recovery by Compression Ratio*)

Based on the RMSE and computational efficiency insights, $M = 40$ and $N = 200$ are selected for this case study, and the corresponding RMSE is about 0.05.

To visualize the effect of the CS approach on CV data collection, a trip made by CV number "2300" is selected. This trip originally has 4,967 speed samples. After applying the CS approach with compression ratio of 0.2 ($M = 40$ and $N = 200$), only 993 samples are retained. Fig. 3a shows part of the original speed samples which are marked in black. Fig. 3b shows the corresponding samples collected with the CS approach which are marked in red. A comparison of Figs. 3a and 3b illustrates that the CS approach efficiently allows CVs to collect and compress the data in real-time.

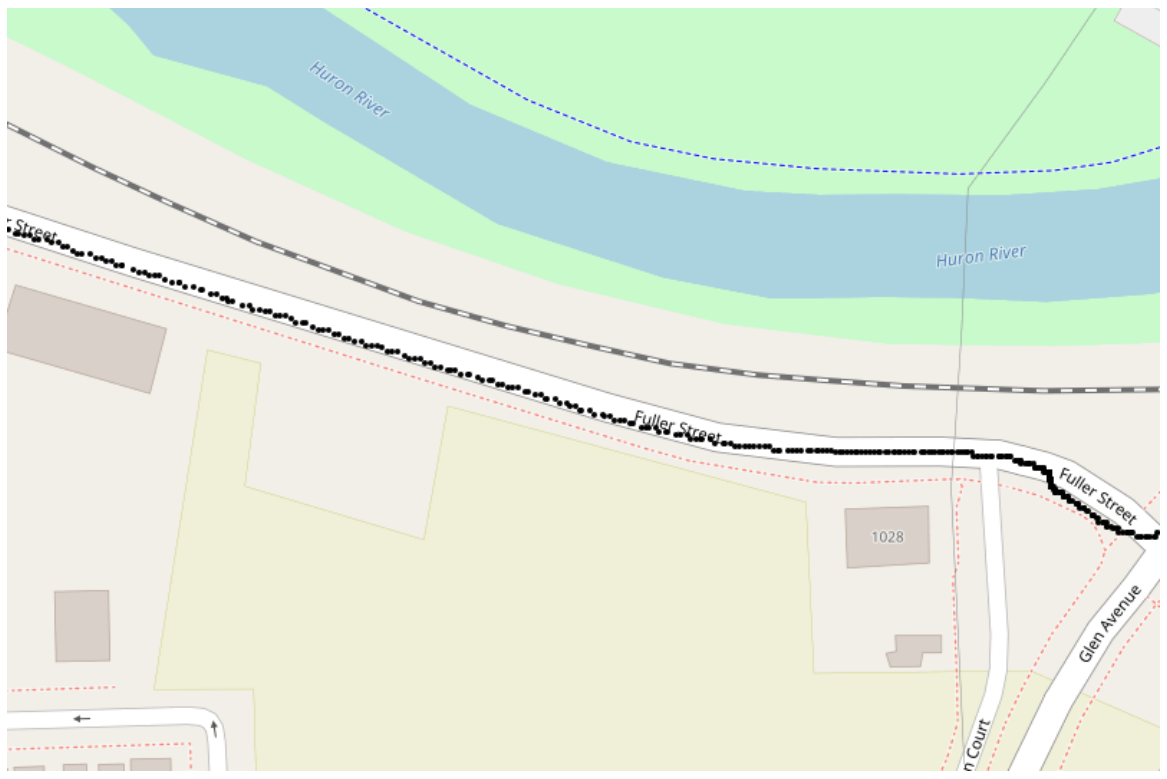

(a)

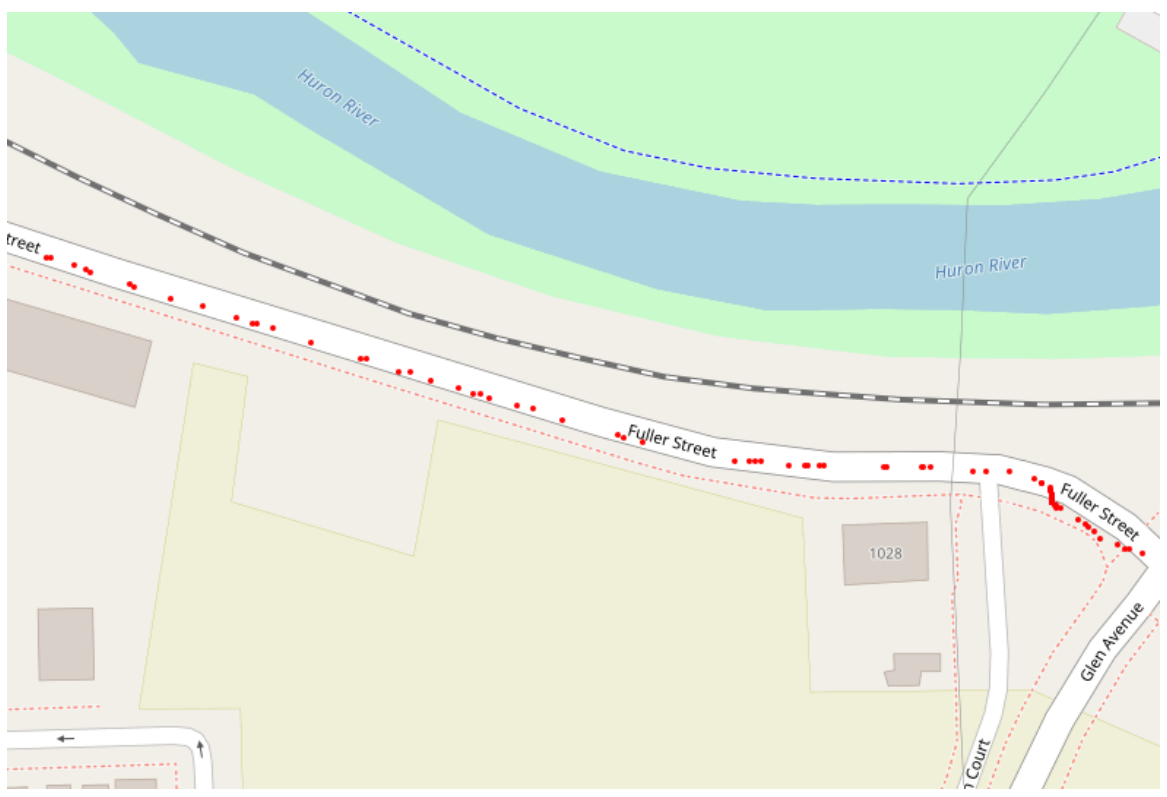

(b)

Fig. 3. Comparison of the Original and Compressed Speed Samples (*a. Original Speed Samples (black); b. Compressed Speed Samples from the CS Approach (red)*)

Fig. 4 further shows the original 4,967 speed samples (black) and the corresponding recovered samples (red). It can be seen that the recovered data are very close to the original samples as the two speed curves overlap with each other, and the RMSE for this trip is only 0.0201.

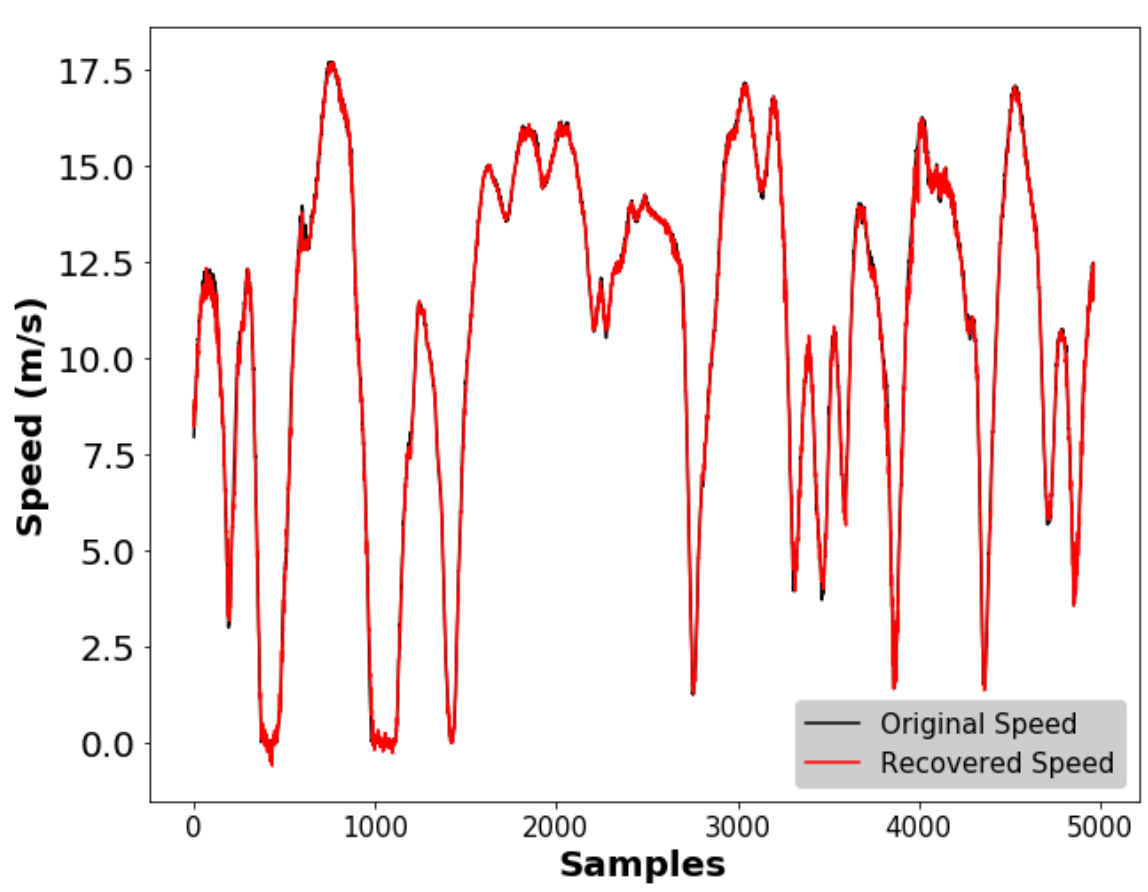

Fig. 4. Comparison of the Original and Recovered Speed Samples

Fig.5 shows the recovery performance of the CS approach by time-of-day. Fig. 5a illustrates that the average speed is the lowest for the morning peak-hour period "7-9". Fig. 5b indicates that this period "7-9" also has the lowest standard deviation of speed. The periods in the afternoon "16-18", "19-21" and "13-15" have the largest standard deviations of speed. The RMSE curve in Fig. 5c indicates that the largest RMSEs occur in time periods "16-18", "19-21" and "13-15". This suggests that the recovery performance of the CS approach may be impacted by whether the driving speed is stable.

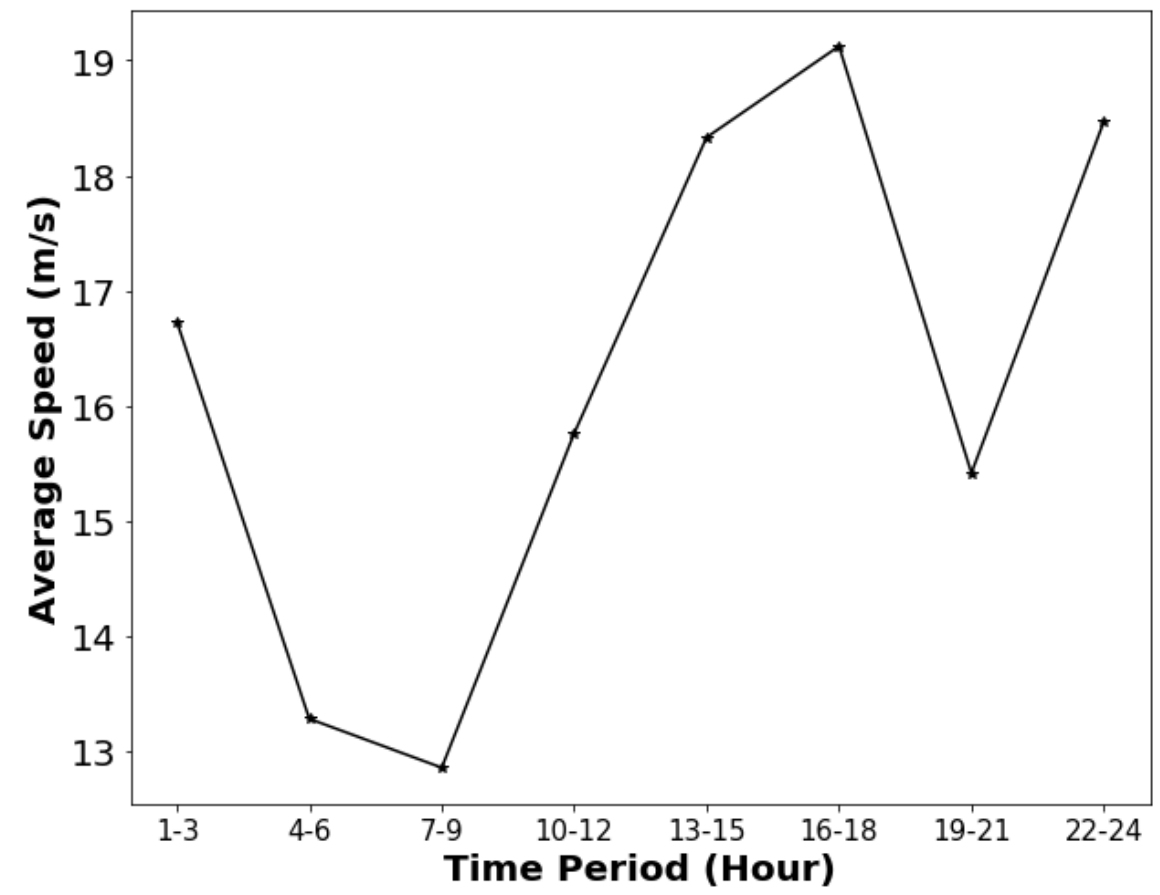

(a)

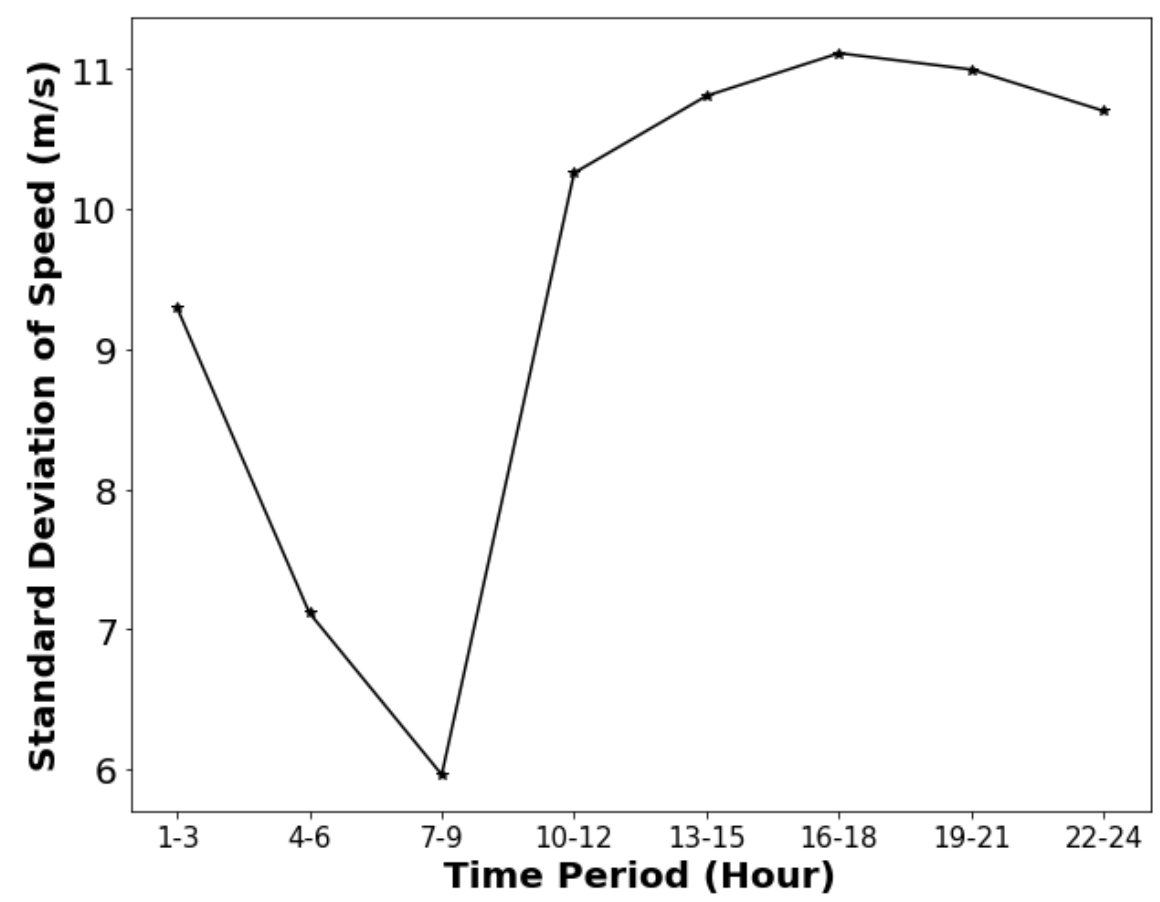

(b)

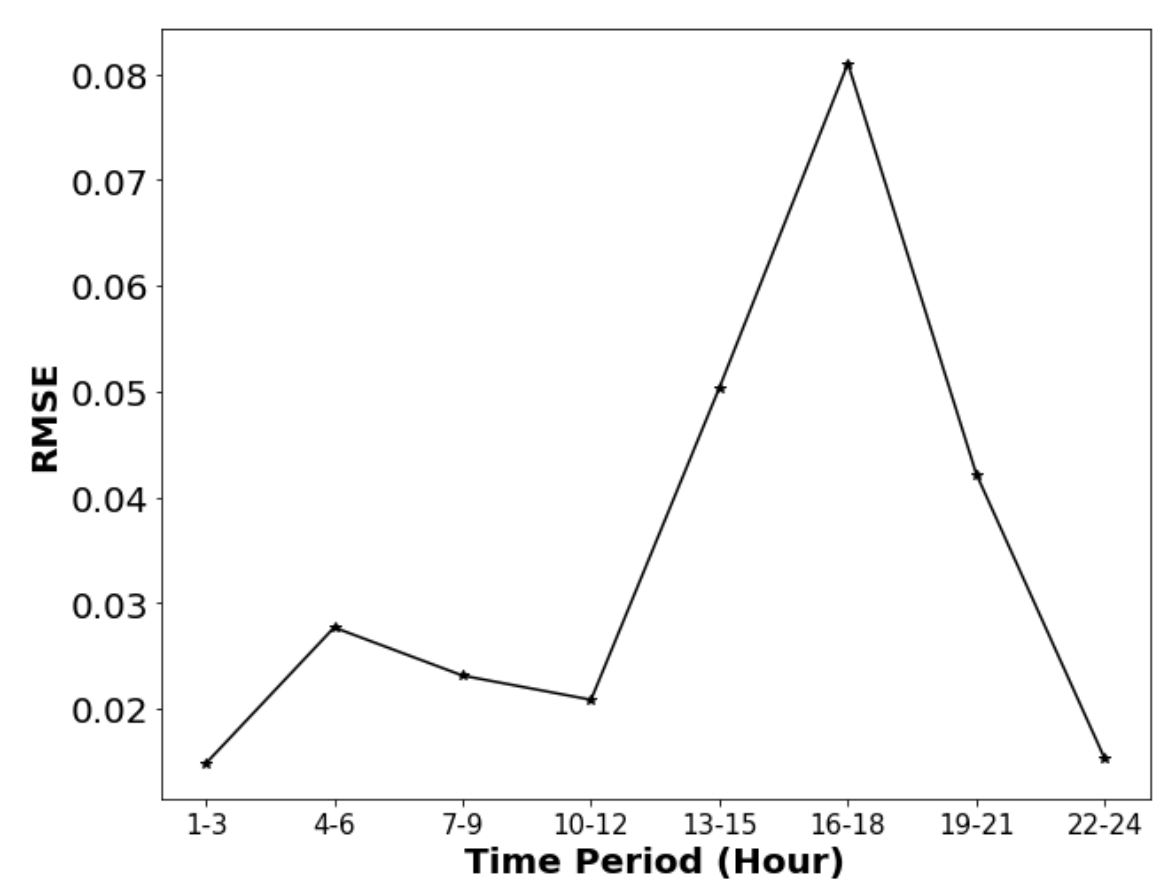

(c)

Fig.5. Recovery Performance of the CS Approach by Time Period (*a. Average Speed by Time Period; b. Standard Deviation of Speed by Time Period; c. RMSE by Time Period*)

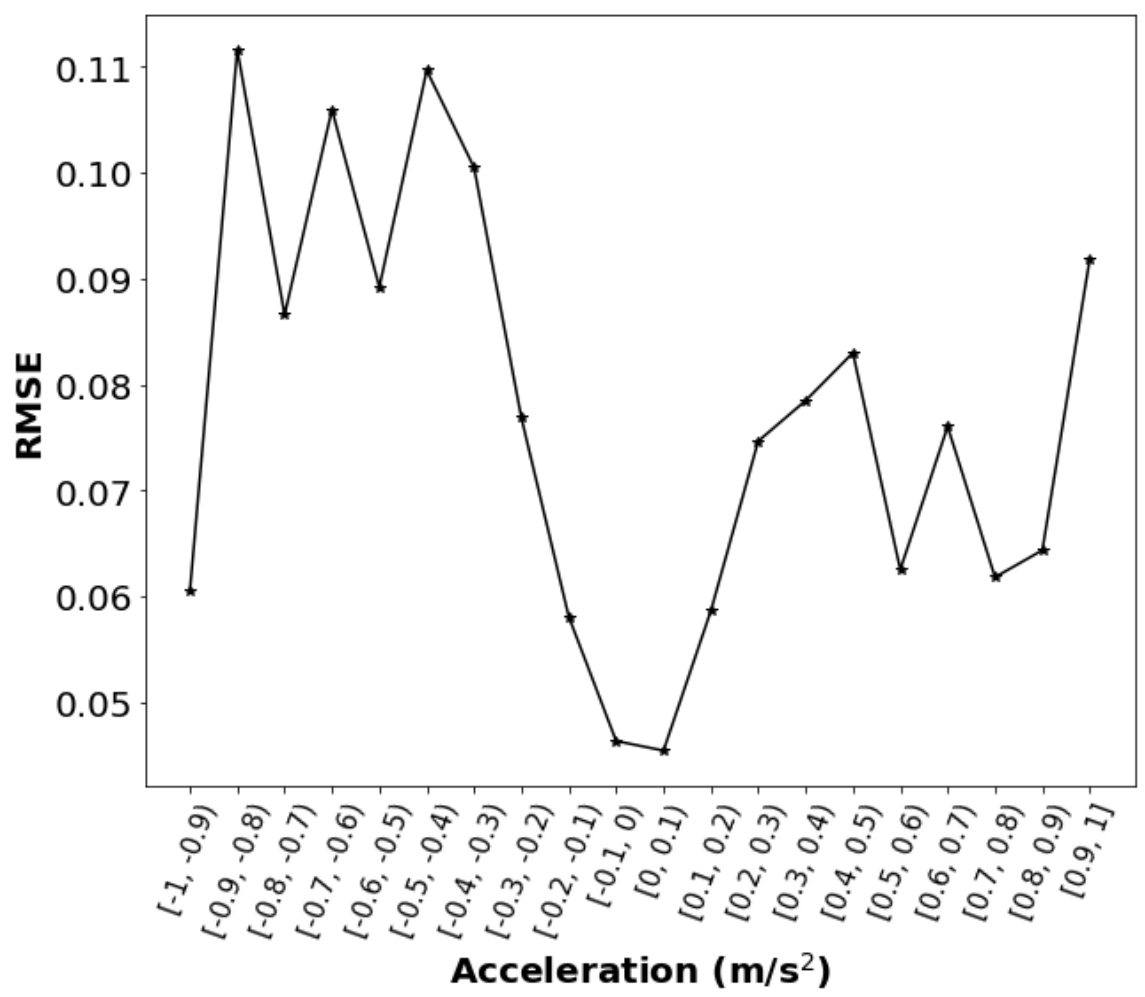

Fig.6. Recovery Performance of the CS Approach by Acceleration

For each speed sample, we can also calculate the corresponding acceleration at that time point. Fig. 6 connects the recovery performance of the CS approach with the acceleration. It can be seen that the RMSEs are very low when the CVs are decelerating or accelerating smoothly (rates in the ranges [-0.1, 0) and [0, 0.1)). For sudden acceleration or deceleration scenarios (for example, when the absolute value of the acceleration is greater than 0.3), the RMSE increases.

## IV. Concluding Comments

With the emergence of CVs, huge amounts of redundant data with little marginal value are being collected, overwhelming storage and communication systems and entailing huge costs. In this paper, we propose a CS approach for efficient CV data collection. It allows CVs to compress the data in real-time such that the original data can be recovered accurately and efficiently when necessary. 10 million BSM speed samples are extracted from the SPMD program. The CS approach is implemented to recapture this dataset with various settings of parameters M and N which determine the compression ratio. It is shown that when the compression ratio is 0.2 ($M = 40$ and $N = 200$), the CS approach can recover the original speed data with RMSE as low as 0.05 and time consumption per recovery close to zero. We further evaluate the recovery performance of the CS approach by time-of-day and acceleration. The results show that the recovery RMSEs of the CS approach are impacted by whether the speeds of CVs are steady or changing smoothly.

In terms of future research directions, we will design a dynamic compression ratio for the CS approach. For example, when CVs are accelerating or decelerating very fast, they should implement a higher compression ratio. We will also test the impact of the CS approach on some transportation management applications, e.g., travel time estimation. With fixed on-board unit capacity, CV data from the CS approach are expected to record more valuable traffic state information that cover longer time periods and longer road segments, leading to more accurate travel times.